\documentclass[11pt,aps,preprint,nofootinbib]{revtex4}
\usepackage[active]{srcltx}
\usepackage[utf8]{inputenc}
\usepackage{latexsym}
\usepackage{amsmath}
\usepackage{graphicx}

\begin{document}

\title{Low-Energy Lorentz Invariance in Lifshitz Nonlinear Sigma Models}

\author{Pedro R. S. Gomes}
\email{pedrogomes@uel.br}
\affiliation{Departamento de F\'isica, Universidade Estadual de Londrina, \\
Caixa Postal 10011, 86057-970, Londrina, PR, Brasil}

\author{M. Gomes}
\email{mgomes@fma.if.usp.br}
\affiliation{Instituto de F\'\i sica, Universidade de S\~ao Paulo\\
Caixa Postal 66318, 05314-970, S\~ao Paulo, SP, Brasil}

%%%%%%%%%%%%%%%%%%%%%%%%%%%%%%%%%%%%%%%%%%%%%%%
\begin{abstract}

This work is dedicated to the study of both large-$N$  and perturbative quantum behaviors of Lifshitz nonlinear sigma models 
with dynamical critical exponent $z=2$ in 2+1 dimensions. We discuss renormalization and  renormalization group aspects
with emphasis on the possibility of emergence of Lorentz invariance at low energies. 
Contrarily to the perturbative expansion, where in general the Lorentz symmetry restoration is delicate and may
depend on stringent fine-tuning, our results provide a more favorable scenario in the large-$N$ framework. 
We also consider supersymmetric extension in this nonrelativistic situation.

\end{abstract}
\maketitle
%%%%%%%%%%%%%%%%%%%%%%%%%%%%%%%%%%%%%%
\section{Introduction}\label{Introduction}

Lorentz invariance is a symmetry of nature, at least at the observable scales of energy \cite{Russell}. 
It is uncertain, however, if it is a property of physics at extremely high energies such as the Planck scale. 
There are some hints that the spacetime itself can be an emergent low-energy concept \cite{Doplicher,Seiberg}, 
so this will be for all its symmetries. 
A lot of attention has been devoted to the study of Lorentz-violating field theories \cite{David}.
An expected feature of any minimal realistic Lorentz-violating model is the manifestation of relativistic invariance  at low energies. 
In this sense, Lorentz symmetry is seen as an emergent phenomenon. 
Although the full spectrum does not exhibit the symmetry it can eventually  appear in specific limits.

This work is dedicated to the study of nonrelativistic theories with emphasis on the Lorentz symmetry restoration at low-energies. 
As a first prototype we consider a Lifshitz-type version of the $O(N)$ nonlinear sigma model which 
admits both large-$N$ and perturbative expansions\footnote{In the relativistic context, a comprehensive account on large-$N$ methods for nonlinear sigma models including the supersymmetric extension is presented in \cite{Justin}.}.
It is characterized by the presence of higher spatial derivative
terms, inducing an anisotropic scaling between space and time weighted by the critical exponent $z$. This kind of deformation
corresponds to a hard breaking of the Lorentz invariance in the sense that the operator content of the theory is modified compared to the
relativistic model  \cite{Pedro}.
At the price of Lorentz symmetry, the presence of higher spatial derivatives shifts the  renormalizability of the model  to higher dimensions compared to its relativistic counterpart. 
In addition to its intrinsic interest as a field theory, this class of nonrelativistic models has an important role in condensed matter 
systems \cite{Fradkin,FradkinBook}. 

To make more precise the type of question we will be mainly concerned along this work, let us take
a simple anisotropic model in $d$ spatial dimensions,
\begin{equation}
S=\int d^dx dt \left(\frac{1}{2}\partial_0\varphi \partial_0\varphi+\cdots-\frac{a_z^2}{2} \varphi \Delta^z\varphi
-\frac{m^{2z}}{2}\varphi^2-\frac{\lambda}{4!}\varphi^4 \right),
\label{I.1}
\end{equation}
where $\Delta\equiv \partial_i \partial_i$ is the spatial Laplacian and the ellipsis refer to spatial derivative terms of order lower than $z$. 
The so-called Lifshitz point is characterized by the vanishing of the mass $m$ and  all terms suggested by the ellipsis. 
At classical level, the action (\ref{I.1}) then  becomes invariant under the scaling 
\begin{equation}
x^0\rightarrow l^z x^0~~~\text{and}~~~x^i\rightarrow l x^i,
\label{l.2}
\end{equation}
at the Lifshitz point in $d=3z$.
To develop a power-counting that naturally takes into account the anisotropic nature of space and time, it
is convenient to assign dimensions for the coordinates as $[x^0]=z$ and $[x^i]=1$ in length units.
The renormalization properties will depend on the values of both $z$ and $d$.
The relativistic case, with $z=1$, is strictly renormalizable in 3+1 spacetime dimensions.
The anisotropic case, with $z=2$, is strictly renormalizable in 6+1 spacetime dimensions~\cite{Gomes}.

Let us take the case $z=2$ in 6+1 spacetime dimensions. The study of Lorentz invariance involves some delicate points.
We cannot naively take the relativistic limit $a_z\rightarrow 0$, since we end up with a nonrenormalizable model
in 6+1 dimensions. In general, the renormalization group $\beta$-functions involve the parameter $a_z$ in the
denominator and  making $a_z\rightarrow 0$ will lead to inconsistencies.
However, we can imagine a sector of energy where the Lorentz symmetry is approximate. In this regime
the parameter $a_z$ flows to small enough values such that it does not interfere in the dynamics
but yet it is nonvanishing. Eventually, this region can be reached only after stringent fine-tuning \cite{Gomes,Iengo}. 
The basic conclusion is that within the
perturbative context, the search for Lorentz invariance is only approximate in the sense that we cannot simply to
turn off the Lorentz violating operators, since we end up with  nonrenormalizable theories.
We shall denote this situation as Lorentz restoration symmetry in the weak form.

A different situation, however, can arise in the framework of the $1/N$ expansion. That is why
the nonlinear sigma model becomes an excellent laboratory to explore that question and enable us to compare different approaches. 
To clarify what we mean, consider the relativistic model (i.e., with $z=1$) in 
the perturbative approach. It is renormalizable in 1+1 dimensions and it becomes nonrenormalizable as
we go to higher dimensions.
However, by performing the $1/N$ expansion, it can be shown that the model is still renormalizable in 2+1 dimensions \cite{Arefeva}.
This feature is translated to the $z=2$ anisotropic version in the following way. The model turns out to be renormalizable in $2+1$, $3+1$, $4+1$ and $5+1$ spacetime dimensions
in the $1/N$ expansion  \cite{Anselmi,Gomes3}. In contrast, we will show that it is perturbatively renormalizable only in 2+1 dimensions (see also   \cite{Farakos}).
There is an overlap situation, namely in 2+1 dimensions, in which even if we take the limit of the coefficients of the
Lorentz violating terms to zero, we still have the possibility of obtaining a renormalizable theory. We will denote this as Lorentz symmetry
restoration in the strong form. This is a special feature which does not occur in the perturbative context. From our calculation
of the renormalization group beta function of the coupling constant, we argue that there is a strong indication that this form of restoration indeed occurs.

In addition to the running of the coupling constant, it is also interesting to determine the flow of the other parameters of the theory.
As we will see, while the renormalization group calculations are hard to overcome in the large-$N$ expansion, most of them can  
be carried out in the perturbative expansion. In turn, the perturbative approach unveils an interesting theory which 
exhibits a $z=2$ Lifshitz scaling at quantum level up to one loop.
 
In the second part of the work these investigations are extended to the supersymmetric case. The construction of supersymmetric models incorporating Lifshitz scaling \cite{Redigolo,Petrov1, Queiruga1,Oz} presents some difficulties mainly when concerned with the superspace formulation.  In that situation, two procedures have been devised accordingly the superalgebra has the standard form or has been deformed by the inclusion of higher spatial derivative terms; in this work we  will discuss both possibilities. Firstly, differently from the 3+1 dimensional case, where the introduction of higher spatial derivatives in the Kählerian 
or in the chiral part of the Wess-Zumino action has some problems \cite{Petrov1}, here the application of the first procedure is straightforward furnishing a consistent field theory what allow us to discuss its quantum properties in the context of large-$N$ expansion.  
In the second approach one has to be careful as the supercharge does not obey
the Leibniz rule. Nevertheless, we show that a consistent model with characteristics analogous to the one resulting from  the stochastic quantization scheme may be constructed \cite{Orlando,Bienzobaz}. A slight connected 
construction of Lifshitz supersymmetry in 2+1 dimensions is presented in \cite{Fujimori}.

This work is organized as follows. In section II, a general analysis of the real and complex Grassmannian models on symmetric space is used to construct the $O(N)$ Lifshitz sigma model with $z=2$. In section III, we discuss the renormalizability of the models for generic values of $z$. In the context of the $1/N$ expansion we obtain the coupling constant beta function for $z=2$ in 2+1 dimensions and analyze the possibility of restoration
of Lorentz symmetry at low energies. In section IV, the perturbative approach is developed to compare and complement the large-$N$ 
studies. In section V, we present a $z=2$ supersymmetric version of the model.
We have included an appendix where we discuss the perturbative renormalization of the Lifshitz nonlinear 
nonlinear sigma model.

%%%%%%%%%%%%%%%%%%%%%%%%%%%%%%%%%%%%%%%%%%%%%%%%%%%%%%%%%

\section{Lifshitz Generalized Sigma Models}

Although we are mainly concerned with the anisotropic $O(N)$ sigma model, in this section we 
discuss a general construction for a special class of
sigma models defined on symmetric spaces, namely the generalized orthogonal 
$O(N)/(O(p)\otimes O(N-p))$ and unitary $U(N)/(U(p)\otimes U(N-p))$ sigma models.
In the relativistic context, this approach provides a unified description of  models exhibiting 
common properties as asymptotic freedom and an infinite number of classical conserved charges in 1+1 dimensions \cite{Brezin,Hikami}.  

Let $g(x)$ a matrix-field that takes values in a simple compact group $G$, governed by an action invariant under global transformations
$g\rightarrow L g R$, where $L,R$ are independent matrices belonging to $G$.  The term involving time derivatives is the usual, 
\begin{equation}
S=\rho \int d^dx dt \text{Tr}\left[ \partial_0 g \partial_0 g^{-1}+\cdots\right],
\label{0.0}
\end{equation}
where $\rho$ is a coupling constant. As the matrix-field $g$ is dimensionless, the dimension of  
the coupling constant reads,
\begin{equation}
[\rho]=z-d.
\label{0.0a}
\end{equation}
For $z=2$ the model is strictly renormalizable in $d=2$ spatial dimensions. 
Marginal operators contain four spatial derivatives and an arbitrary number of $g$'s. Invariance under 
$g\rightarrow L g R$ imposes that we should have an even total number of matrix-fields, with the same number of $g$'s and $g^{-1}$'s. 
Any term involving the product of more than four fields can be reduced, such that we arrive at the most general action
for the generalized anisotropic $z=2$ sigma model containing only marginal operators,
\begin{eqnarray}
S&=&\rho \int d^2x dt \text{Tr}\left[ \partial_0 g \partial_0 g^{-1}+
a_2^2\nabla^2g\nabla^2g^{-1}- a_3 \partial_i g \partial_i g^{-1} \partial_j g \partial_j g^{-1}
-a_4 \partial_i g \partial_j g^{-1} \partial_i g \partial_j g^{-1}\right].
\label{0.1}
\end{eqnarray}
In writing this action we have tacitly assumed that $g^2=1$, which is true for  the class of models that we are interested. 
If this is not the case, of course, the most general action will contain many more terms. 
The dimensionless parameters $a_2, a_3$ and $a_4$ are real and for convenience we set $a_2\geq 0$.
Aiming to study Lorentz symmetry restoration, we include also the relevant operator
\begin{equation}
S_R=\rho\, a_1^2\int d^2x dt \text{Tr} [\partial_i g \partial_i g^{-1}].
\label{0.2}
\end{equation}
The contributions (\ref{0.1}) and (\ref{0.2}) define the $z=2$ Lifshitz version of the generalized sigma model. 

By conveniently restricting
the field matrix $g$ we obtain Lifshitz versions of the usual $O(N)$ and $CP(N)$ sigma models. These models belong to a subclass where
 $g$ varies over a subset of $G$,
such that it can be written as
\begin{equation}
g(x)\equiv q^{-1}(x)g_0 q(x),
\label{0.2a}
\end{equation}
with both $g_0$ and $q$ belonging either to $O(N)$ or $U(N)$ and $g_0^2=1$. This implies $g^2=1$, as stated before. 
For elements $h(x)$ of a subgroup $H$ of $G$ which leaves $g_0$ invariant, i.e., $h^{-1}g_0h=g_0$, note that $g$
is invariant under left multiplication $q(x)\rightarrow h(x)q(x)$. Thus $g$ take values in the coset $G/H$.
Let us consider $G\equiv O(N)$ or $G\equiv U(N)$ and assume a diagonal form for $g_0$,
\begin{equation}
g_0=(\underbrace{1,1,...,1}_{p},\underbrace{-1,-1,...,-1}_{N-p}). 
\label{0.2b}
\end{equation}
Therefore the $g$-field takes values in the coset $\frac{O(N)}{O(p)\otimes O(N-p)}$ or in $\frac{U(N)}{U(p)\otimes U(N-p)}$, 
as $q$ is in $O(N)$ or in $U(N)$.
In these cases, the elements $g_{AB}$ can be written as
\begin{equation}
g_{AB}=2(q^{-1})_{Aa}q_{aB}-\delta_{AB},
\label{0.3}
\end{equation}
with $A,B=1,...,N$ and $a,b=1,...,p$. 
We can now obtain  models involving a set of $N$ scalar fields, $\varphi_a$, invariant under 
$O(N)$ or $U(N)$ transformations, as it follows.
For  $q \in O(N)$,  $q$ is real and $q^{-1}=q^{T}$, the transpose of $q$ whereas for $q \in U(N)$, $q^{-1}=q^{\dagger}$, the Hermitian 
conjugate of $q$. In any case we define
the field $\sqrt{\frac{2g}{N}}\varphi_A^a\equiv q_{aA}$,  where 
$g$ is a new coupling constant. It follows then
\begin{equation}
g_{AB}=\frac{4g}{N}\varphi_A^a\varphi_B^a-\delta_{AB}, 
\label{0.4}
\end{equation}
with $\varphi_A^a \varphi_A^b=\frac{N}{2g}\delta^{ab}$ for $q\in O(N)$ or
\begin{equation}
g_{AB}=\frac{4g}{N}\varphi_A^{a\,*}\varphi_B^a-\delta_{AB}, 
\end{equation}
with $\varphi_A^{a\,*} \varphi_A^b=\frac{N}{2g}\delta^{ab}$ for $q\in U(N)$.

For $p=1$, we get Lifshitz-type versions of  the usual $O(N)/O(N-1)$, or   $U(N)/U(N-1))$. From now on we will restrict ourselves to the case in  which $q\in O(N)$.
The  action (\ref{0.1}) with the contribution of (\ref{0.2}) written in terms of the $\varphi$-fields gives
\begin{eqnarray}
S&=&\int d^2x dt \left[\frac{1}{2}\partial_0\varphi \partial_0\varphi -
\frac{a_1^2}{2}\partial_i\varphi \partial_i\varphi -\frac{a_2^2}{2}\nabla^2\varphi \nabla^2\varphi-\frac{m^4}{2}\varphi^2\right.\nonumber\\
&-&\left.\frac{ g a_3}{N}(\varphi\nabla^2\varphi)(\varphi\nabla^2\varphi)
-\frac{2 g a_4}{N}(\varphi\partial_i\partial_j\varphi)(\varphi\partial_i\partial_j\varphi) -
\frac{\sigma}{\sqrt{2N}}\left(\varphi^2-\frac{N}{2g}\right)\right],
\label{0.5}
\end{eqnarray}
with $\rho$ chosen as $\rho\equiv \frac{N}{32 g}$ and the redefinitions $a_2^2+2a_3\rightarrow a_3$ and 
$a_2^2+a_3+2a_4\rightarrow a_4$. 
We have introduced an auxiliary field to implement the constraint and added a mass term which is innocuous due to the constraint.
The sum over the index $A$ is implicit. Note the presence of quartic terms which, as we shall see, are
necessary to ensure renormalizability in $d=2$. %\cite{Gomes3}. 

%%%%%%%%%%%%%%%%%%%%%%%%%%%%%%%%%%%%%%%%%%%%%%%%%%%%

\section{Large-$N$ Expansion}\label{LN}

As argued in Sec. \ref{Introduction}, we will study the quantum behavior of the theory (\ref{0.5}) in the framework of the
$1/N$ expansion. We start by discussing its renormalizability 
and then proceed with a renormalization 
group analysis. As we shall describe,  
the general procedure to cope with divergences in the case of $d=2$ spatial dimensions can be understood by constructing the 
possible counterterms.  

\begin{figure}[!h]
\centering
\includegraphics[scale=0.6]{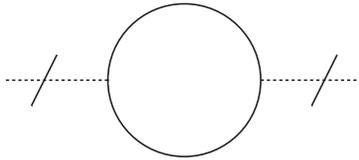}
\caption{Dotted lines represent $\sigma$-field whereas continuous lines represent $\varphi$-field.
The $\sigma$-propagator, $\Delta_{\sigma}$, is the inverse of this one-loop amputated diagram.}
\label{Fig1}
\end{figure}

The propagator for the $\varphi_A$-fields is 
\begin{equation}
\Delta_{AB}(p)=\frac{i \delta_{AB}}{p_0^2-a_1^2{\bf p}^2-a_2^2({\bf p}^2)^2-m^4+i\epsilon}.
\label{4.1}
\end{equation}
Although there is no kinetic term for the $\sigma$-field in the classical action (\ref{0.5}), it is
dynamically generated, enabling us to define its propagator in terms of the inverse of the analytic expression for the diagram in Fig. \ref{Fig1},
\begin{equation}
-\Delta^{-1}_{\sigma}(p)=\int \frac{dk_0}{2\pi}\frac{d^2k}{(2\pi)^d} \frac{i}{(k_0+p_0)^2-\omega_{k+p}^2+i\epsilon}\,
 \frac{i}{k_0^2-\omega_{k}^2+i\epsilon},
 \label{4.2}
\end{equation}
where we introduced the nonrelativistic frequency $\omega_k^2\equiv a_1^2 {\bf k}^2+a_2^2 ({\bf k}^2)^2+m^4$.
The generalization for arbitrary values of $z$ is obtained by including higher powers of momenta in the frequency, namely,  
$\omega_k^2\rightarrow a_1^2 {\bf k}^2+a_2^2 ({\bf k}^2)^2+\cdots+ a_z^2({\bf k}^2)^z+m^4$.

For arbitrary $z$ and $d$, the inverse of the propagator in Eq. (\ref{4.2}) behaves as $|{\bf p}|^{d-3z}$.   
The integral  converges for $d-3z<0$, which we shall impose from now on.
For a generic diagram $G$ with $L$ loops, $n_{\varphi}$ internal lines of $\varphi$ and $n_{\sigma}$ internal 
lines of $\sigma$, the superficial degree of divergence is
\begin{equation}
d(G)= (z+d)L-2zn_{\varphi}+(3z-d)n_{\sigma}.
\end{equation}
By using the Euler relation $L=n_{\varphi}+n_{\sigma}-V+1$, with $V$ being the number of vertices, 
and the topological relations $2n_{\varphi}+N_{\varphi}=2V$ and $2n_{\sigma}+N_{\sigma}=V$, with 
$N_{\varphi}$ and $N_{\sigma}$ being the number of external lines of the corresponding fields, we get
\begin{equation}
d(G)=z+d-\frac{(d-z)}{2}N_{\varphi}-2zN_{\sigma}.
\label{4.3}
\end{equation}
It is not desirable to have the degree of divergence increasing with the number of external lines of $\varphi$. Thus we shall impose 
$d\geq z$, arriving to a criterion of renormalizability: $z\leq d <3z$. 
For $d=z=2$, the superficial degree of divergence (\ref{4.3}) reduces to
\begin{equation}
d(G)=4-4N_{\sigma}.
\label{4.4}
\end{equation}
Note that it does not depend on the number of external lines involving $\varphi$. Thus all 
1PI\footnote{We mean 1PI functions only with respect to $\varphi$-lines, not with respect to $\sigma$-lines \cite{Gomes3}.} 
functions involving only
external $\varphi$-fields are quartically divergent.  There is, however, a general mechanism for
cancellation of divergences, which can be understood in view of the impossibility to construct arbitrary counterterms because of the constraint $\varphi^2=N/2g$.
We will consider firstly 1PI diagrams with no $\sigma$-type external lines, i.e., $N_{\sigma}=0$.

$\bullet$ For $N_{\varphi}=2$, the quartic divergences can be eliminated
by means of a renormalization of the parameters $a_1$, $a_2$ and $m$, in addition to the wave function
renormalization. The counterterms are of the form
\begin{equation}
\partial_0\varphi\partial_0\varphi,~~~\partial_i\varphi\partial_i\varphi,~~~\text{and}~~~\nabla^2\varphi \nabla^2\varphi.
\label{4.5}
\end{equation}

$\bullet$ For $N_{\varphi}=4$. The only nonreducible counterterms are,
\begin{equation}
(\varphi \partial_i \partial_j\varphi) ( \varphi \partial_i\partial_j\varphi)~~~\text{and}  ~~~(\varphi \nabla^2\varphi) ( \varphi \nabla^2\varphi).
\label{4.6}
\end{equation}
For example, if the derivatives are acting only on one pair of fields with contracted indices it can be reduced to a
two point function as $(\nabla^2\varphi \nabla^2\varphi) ( \varphi \varphi)\sim (\nabla^2\varphi \nabla^2\varphi) $
due to the constraint. When we have only two derivatives, or it is zero, $(\varphi \partial_{\mu} \varphi)(\varphi \partial_{\nu} \varphi) $
since $(\varphi \partial_{\mu} \varphi)\sim \partial_{\mu}\varphi^2$, or it is reducible to the two point function
$(\varphi \partial_{\mu}\partial_{\nu} \varphi)(\varphi  \varphi) \sim \varphi \partial_{\mu}\partial_{\nu} \varphi$.
Terms with no derivatives are always reducible.

$\bullet$ For $N_{\varphi}> 4$. In this case, all 1PI diagrams are reducible to the previous cases.
For example, consider the nonzero counterterm with four spatial derivatives,
\begin{equation}
(\varphi \nabla^2\varphi) ( \varphi \nabla^2\varphi) ( \varphi \varphi) ( \varphi \varphi) \cdots
\sim (\varphi \nabla^2\varphi) ( \varphi \nabla^2\varphi).
\label{4.7}
\end{equation}
Now let us turn on to the case $N_{\sigma}=1$. Now all 1PI involving external $\varphi$-lines are
logarithmically divergent.

$\bullet$ For $N_{\varphi}=0$, the counterterm is proportional to $\sigma$.

$\bullet$ For $N_{\varphi}>0$ all counterterms are reducible to the previous case since
$\sigma(\varphi\varphi)...(\varphi\varphi)\sim \sigma$.

So all counterterms necessary to absorb the divergences are of the form of the original Lagrangian showing that the model 
is renormalizable. At level of 1PI functions, the divergence canceling mechanism is that for any 1PI diagram we can associate a set of
diagrams, at the same order in $1/N$, such that the sum of all diagrams will be finite \cite{Gomes3}.

\begin{figure}[!h]
\centering
\includegraphics[scale=0.8]{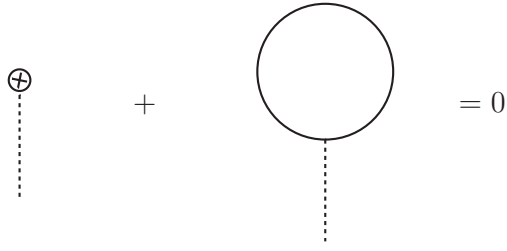}
\caption{Gap equation. To the crossed vertex it is associated the factor $\frac{i}{2g}\sqrt{\frac{N}{2}}$.}
\label{Fig2}
\end{figure}

%%%%%%%%%%%%%%%%%%%%%%%%%%%%%%%%%%%%%%%%%%%%%%%%%%%
\subsection{Running Couplings and Lorentz Resurgence} 

To investigate the low-energy limit we need to determine the running couplings of the theory
under the renormalization group flow. Unfortunately, most of calculations involving nonrelativistic propagators 
like (\ref{4.1}) and (\ref{4.2}) are 
hard enough standing out of our abilities. In contrast, the renormalization group beta function of the coupling 
constant $g$ can be easily obtained from the gap equation. 

The gap equation, diagrammatically shown
in Fig. \ref{Fig2}, follows from the requirement that the auxiliary field has a vanishing expectation value, 
$\langle\sigma\rangle=0$. The corresponding expression is
\begin{equation}
\frac{1}{2g}=\int\frac{dk_0}{2\pi}\frac{d^2k}{(2\pi)^2}\frac{i}{k_0^2-\omega_k^2(m)}.
\label{3.1}
\end{equation}
We are making explicit the mass dependence in the frequency and omitting the $i\epsilon$ prescription. The logarithmic divergence can be eliminated by adopting a Pauli-Villars regularization
\begin{equation}
\frac{1}{2g(\Lambda)}=\int\frac{dk_0}{2\pi}\frac{d^2k}{(2\pi)^2}\left[\frac{i}{k_0^2-\omega_k^2(m)}-\frac{i}{k_0^2-\omega_k^2(\Lambda)}\right],
\label{3.2}
\end{equation}
where $\Lambda$ is an ultraviolet cutoff. To proceed we firstly calculate the $k_0$-integral in the complex plane
and then over ${\bf k}$. We obtain
\begin{equation}
\frac{1}{2g(\Lambda)}=-\frac{1}{8 \pi a_2}\ln \left[\frac{a_1^2+2 a_2 m^2}{a_1^2+2a_2\Lambda^2}\right].
\label{3.3}
\end{equation}
The cutoff can be eliminated by introducing the renormalized coupling constant $g_R(\mu)$ at the scale $\mu$ according to
\begin{equation}
\frac{1}{2g_R(\mu)}=\frac{1}{2g(\Lambda)}+\frac{1}{8 \pi a_2}\ln \left[\frac{a_1^2+2 a_2 \mu^2}{a_1^2+2a_2\Lambda^2}\right].
\label{3.4}
\end{equation}
Plugging this relation into Eq. (\ref{3.3}), we get
\begin{equation}
\frac{1}{2g_R(\mu)}=\frac{1}{8 \pi a_2}\ln \left[\frac{a_1^2+2 a_2 \mu^2}{a_1^2+2a_2m^2}\right].
\label{3.5}
\end{equation}
The renormalization group beta function, $\beta_{g_R}\equiv\mu \frac{\partial g_R}{\partial \mu}$, can be easily calculated
\begin{equation}
\beta_{g_R}=-\frac{1}{\pi}\frac{\mu^2}{(a_1^2+2a_2\mu^2)}g_R^2,
\label{3.6}
\end{equation}
showing that the theory is asymptotically free (remember that we set $a_2\geq0$). 
This result exhibits interesting features which give
some clues that Lorentz resurgence in the strong form is possible at low energies.
Note that the beta function is well defined in both situations i) $a_1=0$ and $a_2\neq 0$ and
ii) $a_1\neq 0$ and  $a_2= 0$. Of course, the theory is meaningless if $a_1=0$ and $a_2= 0$ simultaneously. 
In the case $a_1=0$ with $a_2\neq 0$ the beta function reduces to
\begin{equation}
\beta_{g_R}=-\frac{1}{2\pi a_2}g_R^2,
\label{3.7}
\end{equation}
with only a trivial fixed point. On the other hand, for $a_2=0$ with $a_1\neq 0$, we get
\begin{equation}
\beta_{g_R}=-\frac{1}{\pi}\frac{\mu^2}{a_1^2}g_R^2.
\label{3.8}
\end{equation}
In this case it is convenient to introduce a dimensionless coupling constant, $\bar{g}_R\equiv \mu^2 g_R$,
such that the beta function becomes
\begin{equation}
\beta_{\bar{g}_R}=-\frac{1}{\pi a_1^2} \bar{g}_R(\bar{g}_R-\bar{g}^*_R),
\label{3.9}
\end{equation}
where $\bar{g}^*_R\equiv 2\pi a_1^2$ is the nontrivial fixed point, coinciding with the relativistic case in 2+1 spacetime dimensions 
in a system of units such that the light speed is $a_1$ \cite{Arefeva}. 

To make precise the issue of Lorentz restoration at low energies,
we can divide the parameters in two groups according to their expected behavior under renormalization 
group flow: i) $a_1$ and $g$; these parameters should be stable 
at low-energies, i.e., they must be nonvanishing, and 
ii) the Lorentz-breaking parameters $a_2$, $a_3$, and $a_4$; these parameters should vanish smoothly at 
low-energies.
The emergence of Lorentz symmetry at low-energies in the strong form is based on the possibility 
of keeping renormalizability even when $a_2\rightarrow 0$ with $a_1\neq 0$. 
The beta function (\ref{3.6}) has a behavior compatible with the strong form, 
since it is stable in the limit $a_2\rightarrow 0$. 

The unanswered question is if the parameters $a_2$, $a_3$, and $a_4$ flow to zero at low energies.
Unfortunately we are not able to overcome the difficulties with the loop calculations involving nonrelativistic propagators, 
but this is expected in view of the fact that the relativistic limit there exists. 
If we imagine a common physical origin for the breaking of the Lorentz symmetry at higher energies, it is natural 
to assume that the behavior of $a_3$ and $a_4$ are tied to the behavior of $a_2$, i.e., they vanish as $a_2\rightarrow 0$.
Formally this can be done by means of the Zimmerman's method of reduction of coupling constants \cite{Zimmermann}. In this 
situation, $a_2$ can be thought as the fundamental Lorentz breaking parameter and $a_3$ and $a_4$ 
are induced as counterterms.
This will imply relations between the corresponding beta functions
\begin{equation}
\beta_{a_2}\frac{d a_3}{d a_2}=\beta_{a_3}~~~\text{and}~~~\beta_{a_2}\frac{d a_4}{d a_2}=\beta_{a_4}.
\end{equation}
Thus if $a_2\rightarrow 0$ at low energies, so do $a_3$ and $a_4$.

%%%%%%%%%%%%%%%%%%%%%%%%%%%%%%%%%%%%%%%%%%%%%%%%%%%

\section{Perturbative Expansion}

The large-$N$ expansion provides a favorable framework concerning Lorentz symmetry restoration since the $\beta_g$-function is 
stable even when $a_2\rightarrow 0$ with $a_1\neq 0$. Our goal now is to investigate the perturbative 
expansion to compare and complement the large-$N$ results. The discussion of perturbative renormalization at all orders is left to 
the appendix \ref{BB}.

By decomposing the fields as $\phi\equiv\varphi_{1}$ and $\pi_{A}\equiv \varphi_{A} $, for $A\not =1$, and replacing 
\begin{equation}
 \phi =\sqrt{\frac{N}{2g}-\pi^{2}} =\left(\frac{N}{2g}\right)^{1/2}- \left(\frac{g}{2N}\right)^{1/2}\pi^{2}-\left(\frac{g}{2N}\right)^{3/2}(\pi^{2})^{2}+\cdots,
 \label{00.7}
 \end{equation}
into Eq. (\ref{0.5}), we obtain up to quartic terms in $\pi$,
\begin{eqnarray}
S&=&\int d^2x dt \left[\frac{1}{2}\tilde{\partial_\mu}\pi\tilde{ \partial^\mu}\pi +
\frac{g}{N}(\pi\tilde{\partial_{\mu}}\pi)^{2}-\frac{a_2^2}{2}\nabla^2\pi \nabla^2\pi-\frac{2 g a_4}{N}(\partial_{i}\pi\partial_j\pi)^{2} \right.\nonumber\\
&-&\left.\frac{g a_2^2}{N}(\pi\nabla^2\pi)^{2}
-\frac{g (a_{2}^{2}+a_{3})}{N}(\partial_{i}\pi\partial_{i}\pi)^{2}-\frac{2g a_{2}^{2}}{N}(\partial_{i}\pi\partial_{i}\pi)(\pi\nabla^{2}\pi)+\cdots\right],
\label{0.6}
\end{eqnarray}
where we have defined $\tilde{\partial_{\mu}}\equiv (\partial_{0}, a_{1}\partial_{i})$ and 
$\tilde{\partial^{\mu}}\equiv (\partial_{0},- a_{1}\partial_{i})$. 
Contrarily to the large-$N$ expansion, there is no mass generation here and we need to deal with infrared divergences. 
To keep them under control we include into the action the coupling to a constant external field $h$,
 \begin{equation}
 h\phi = h\left[\left(\frac{N}{2g}\right)^{1/2}- \left(\frac{g}{2N}\right)^{1/2}\pi^{2}-\left(\frac{g}{2N}\right)^{3/2}(\pi^{2})^{2}+\cdots
 \right],
 \label{0.7a}
 \end{equation}
which furnishes the mass to the $\pi$-field, namely,  
its propagator is the same as (\ref{4.1}) with $m^4\equiv h\left(\frac{2g}{N}\right)^{1/2}$. The complete action is then 
\begin{equation}
S\rightarrow S+ \int d^2x dt\,h\phi.
\label{0.7b}
\end{equation}
\begin{figure}[!h]
\centering
\includegraphics[scale=0.6]{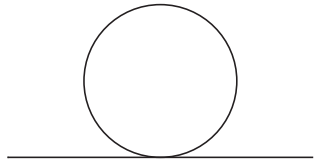}
\caption{One-loop contribution to the two-point function for the $\pi$-field.}
\label{Fig3}
\end{figure}

It is of primary interest to understand the behavior of the coupling constant $g$ as well as the
Lorentz violation parameters $a_2$, $a_3$ and $a_4$ under the renormalization group flow. 
By employing the same sort of reasoning used in the large-$N$ expansion, we can think of $a_3$ and $a_4$ induced from $a_2$.  
To fix them we calculate the one-loop divergent parts to the 1PI two-point function $\langle\pi(p)\pi(-p)\rangle$ of Fig. \ref{Fig3}.
We have several contributions coming from different vertices  present in (\ref{0.6}) and (\ref{0.7a}). They are: 
 
$\bullet$ From the vertex $-\frac{2 g a_4}{N}(\partial_{i}\pi\partial_j\pi)^{2}$:
\begin{equation}
-\frac{ig a_{1}^{2}a_{4}(2N+4){\bf p}^{2}}{4\pi a_{2}^{3}N(d-2)}.
\label{0.8a}
\end{equation}

$\bullet$  From the vertex $-\frac{g a_2^2}{N}(\pi\nabla^2\pi)^{2}$:
\begin{equation}
\frac{ig a_{2}({\bf p}^{2})^2}{2\pi N(d-2)}-\frac{iga_{1}^{2}{\bf p}^{2}}{2\pi a_{2}(d-2)}-\frac{ig(-3a_{1}^{4}+4a_{2}^{2}m^{4})}{16 \pi a_{2}^{3}N(d-2)}.
\label{0.9a}
\end{equation}

$\bullet$ From the vertex $-\frac{g(a_{2}^{2}+a_{3})}{N}(\partial_{i}\pi\partial_{i}\pi)^{2}$:
\begin{equation}
-\frac{ig a_{1}^{2}(a_{2}^{2}+a_{3}){\bf p}^{2}}{2 \pi a_{2}^{3}(d-2)}.
\label{0.10a}
\end{equation}

$\bullet$ From the vertex $-\frac{2ga_{2}^{2}}{N}(\partial_{i}\pi\partial_{i}\pi)(\pi\nabla^{2}\pi)$:
\begin{equation}
i\frac{a_{1}^{2}g(N-1) {\bf p}^{2}}{\pi a_{2}N(d-2)}.
\label{0.11a}
\end{equation}

$\bullet$ From the usual relativistic vertex (i.e., the term $g/N (\pi\tilde{\partial_{\mu}}\pi)^{2}$):
\begin{equation}
-\frac{i\tilde p^2 g}{2\pi a_2N(d-2)}-\frac{ig(-a_{1}^4+4 a_{2}^2 m^4)}{16 \pi a_{2}^{3}N  (d-2)}-\frac{ia_{1}^{4}g}{4 \pi a_{2}^{3} N  (d-2)},
\label{0.12a}
\end{equation}
with $\tilde{p}^2\equiv p_0^2-a_1^2{\bf p}^2$.

$\bullet$ From the vertex $(\pi^{2})^{2}$:
\begin{equation}
\left(\frac{g}{2N}\right)^{3/2}\frac{ih(N+1)}{\pi a_{2}(d-2)}=\frac{i g m^4(N+1)}{4\pi a_{2}N(d-2)}.
\label{0.14a}
\end{equation}

By collecting the above results, the dimensionally regularized, but yet unrenormalized, two-point vertex function up to one-loop order is 
\begin{eqnarray}
\Gamma^{(2)}_{u}(p)&=& i \left[{\tilde p}^{2}-a_{2}^{2}({\bf p}^{2})^{2}-m^{4}-\frac{ p_{0}^2 g}{2\pi a_2 N(d-2)}+
\frac{g a_{2}({\bf p}^{2})^2}{2\pi N(d-2)}\right.\nonumber\\
& -& \left.\frac{ga_{1}^{2}\left[(2a_{4}+a_{2}^{2})\frac{1}{N}+a_{3}+a_{4}\right] {\bf p}^{2}}{2\pi a_{2}^{3}(d-2)}+
\frac{gm^4(N-1)}{4 \pi a_{2}N(d-2)}\right]+\mbox{finite contributions}.
\label{0.8}
\end{eqnarray}
The renormalization of the model follows by performing the reparametrization of the Lagrangian according to,
\begin{equation}
\pi_{A} =  Z^{1/2}\pi_{A\, R },~~~ g =\frac{Z_{g}}{Z^{2}}g_{R}, ~~~a^{2}_{1,2}=\frac{Z_{1,2}}{Z}a^{2}_{1,2\,R},
~~~a_{3,4}=\frac{Z_{3,4}}{Z}a_{3,4\,R}~~~\text{and}~~~
h=\frac{h_{R}}{Z^{1/2}}.
\label{0.8c}
\end{equation}
Redefining also the renormalized mass by
$m_{R}^{4}=h_{R}\left(\frac{2g_{R}}{N}\right)^{1/2}$, we have $m^{4}=\left(\frac{Z_{g}^{1/2}}{Z^{3/2}}\right)m_{R}^{4}$.
Thus the  action (\ref{0.6}) together with the contributions from the term $h\phi$, written  in terms of renormalized quantities has the form
\begin{eqnarray}
S&=&\int d^2x dt \left[\frac{Z}{2}{\partial_0}\pi{ \partial_0}\pi -\frac{a_{1\,R}^{2}Z_{1}}{2}{\partial_i}\pi{ \partial_i}\pi -
\left(\frac{Z_{g}}{Z}\right)^{1/2}m^{4}_{R}\frac{\pi^{2}}2+
\frac{g_{R}Z_g}{N}(\pi\tilde{\partial_{\mu}}\pi)^{2}\right.\nonumber\\
&-&\frac{a_{2\,R}^2Z_{2}}{2}\nabla^2\pi \nabla^2\pi-\frac{g_{R}Z_g}{ZN} \left[2 a_{4 R}^{2}Z_{4} (\partial_i\pi\partial_j\pi)^{2} + a_{2 R}^{2}Z_{2}(\pi\nabla^2\pi)^{2}
+\right.\nonumber\\
&+&( a_{2 R}^{2}Z_{2}+a_{3 R}^{2}Z_{3})(\partial_{i}\pi\partial_{i}\pi)^{2}+ 2 a_{2 R}^{2}Z_{2}(\partial_{i}\pi\partial_{i}\pi)(\pi\nabla^{2}\pi)]\Big{]}.
\label{0.7}
\end{eqnarray}
To avoid heavy notation we omit the subscript $R$  from the renormalized $\pi$-fields. 
By taking into account (\ref{0.8}), we determine the counterterms
\begin{equation}
Z=1+ \frac{g_{R}}{2\pi a_{2 R}N(d-2)}, ~~~ Z_{1}=1-\frac{(a_{4R}^{2}+a_{2R}^{2})(N+1)g_{R}}{\pi a_{2R}^{3}N(d-2)} ~~~\text{and}~~~ 
Z_2= 1+\frac{g_{R}}{2\pi a_{2 R}N(d-2)},
\end{equation}
with $Z_{g}$ fixed from the renormalization of the mass term 
\begin{equation}
m^4 \pi^{2}=\left(\frac{Z_{g}}{Z}\right)^{1/2}m_{R}^{4}\pi_R^{2}\qquad \rightarrow\qquad  \frac{Z_{g}}{Z}= 1+\frac{(N-1)g_{R}}{2\pi a_{2R}N(d-2)},
\end{equation} 
so that 
\begin{equation}
Z_{g}=1+\frac{g_{R}}{2 a_{2R}\pi(d-2)}.  
\label{0.20}
\end{equation}
The unrenormalized and renormalized vertex functions are related by
\begin{equation}
\Gamma_R^{(N_{\pi})}(p, m_{R},g_{R}, a_{iR}, \mu)= Z^{\frac{N_{\pi}}{2}}\Gamma^{(N_{\pi})}(p,m,g,a_i),~~~i=1,...,4,
\label{0.21}
\end{equation}
where $p$ denotes the external momenta and $N_{\pi}$ is the number of external lines of $\pi$-field. By imposing 
\begin{equation}
0=\mu\frac{d\phantom a}{d\mu} \Gamma^{(N_{\pi})}(p,m,g,a_i),
\label{0.22}
\end{equation}
we obtain the renormalization group equation
\begin{equation}
\left[\mu \frac{\partial\phantom a}{\partial\mu}+ \beta_{g_{R}}\frac{\partial\phantom a}{\partial g_{R}}+ \sum_{i=1}^{4}\beta_{a_{iR}}\frac{\partial\phantom a}{\partial a_{iR}} - \gamma\left(\frac{N_{\pi}}{2}-\frac{h_{R}}2\frac{\partial\phantom a}{\partial h_{R}}\right)\right]\Gamma_R^{(N_{\pi})}=0.
\label{0.23}
\end{equation}
By introducing the dimensionless coupling constant through the replacement $g_R\rightarrow \mu^{2-d}g_R$, we can use (\ref{0.23})
to determine the renormalization group functions. 
By taking $N_{\pi}=2$ in (\ref{0.23}) and comparing the corresponding coefficients 
of $m_R^4$, $p_0^2$, ${\bf p}^2$ and $({\bf p}^2)^2$, we find
\begin{equation}
\gamma=\frac{g_R}{2\pi a_{2R} N},~~~\beta_{g_R}=-\frac{ g_{R}^{2}}{2\pi a_{2 R}}\left(\frac{N-2}{N}\right),~~~\beta_{a_{2R}}=0,
\label{0.24}
\end{equation} 
and
\begin{equation}
\beta_{a_1}=\frac{a_{1R}\gamma}{2}+\frac{a_{1R}g_R}{4\pi a_{2R}^3}\left[(2a_{4R}+a_{2R}^2)\frac{1}{N}+a_{3R}+a_{4R}\right].
\label{0.25}
\end{equation}
Note that the $\beta_g$-function is not well-behaved in the limit $a_2\rightarrow 0$, 
which is a sign of nonrenormalizability when we turn off the Lorentz-violating parameter $a_2$. 
Thus in the perturbative context the Lorentz invariance cannot be realized in the strong form. 
Note also that $\beta_g$ coincides with (\ref{3.7}) in the large-$N$ limit. 
As in the large-$N$ expansion, it is natural to expect that $a_3= O(a_2 g)$ and $a_4= O(a_2 g)$, since we can 
think in terms of the reduction of coupling constants. In this case, at leading order, 
\begin{equation}
\beta_{a_{1R}}\approx \frac{a_{1R}g_R}{2\pi a_{2R}N}.
\label{0.26}
\end{equation}
With this we can easily to solve the renormalization group equations 
\begin{equation}
g_R(\mu)=\frac{g_0}{1+\frac{g_0}{2\pi a_{2R}}\left(\frac{N-2}{N}\right)\ln\frac{\mu}{\mu_0}},
\label{0.27}
\end{equation}
where $\mu_0$ is the scale which defines the initial value of parameters, for example, $g_R(\mu_0)\equiv g_0$. 
Using this in (\ref{0.26}), we find
\begin{equation}
a_{1R}(\mu)=a_{1R}(\mu_0)\left[1+\frac{g_0}{2\pi a_{2R}}\left(\frac{N-2}{N}\right)\ln\frac{\mu}{\mu_0} \right]^{\frac{1}{N-2}},
\label{0.28}
\end{equation}
showing that $a_1$ goes monotonically to zero at low energies until to reach the infrared critical value 
$\mu_{IR}=\mu_0 \exp\left(-\frac{2\pi a_{2R}}{g_0}\frac{N}{N-2}\right)$.

%%%%%%%%%%%%%%%%%%%%%%%%%%%%%%%%%%%%%%%%%%%%%%%

\subsection{Anisotropic $z=2$ scale invariance at quantum level?}

It is interesting to note that for $N=2$ the beta function vanishes. 
In the relativistic case, the beta function also vanishes for $N=2$, but this is just a reflex that 
the relativistic nonlinear sigma model becomes trivial when $N=2$. By parameterizing the fields as 
\begin{equation}
(\varphi_1,\varphi_2)\equiv\frac{1}{\sqrt{g}}(\cos\theta,\sin\theta), 
\label{si1}
\end{equation}
which automatically solve the constraint, the action reduces to a free theory
\begin{equation}
\mathcal{L}=\frac{1}{2}\partial_{\mu}\theta\partial^{\mu}\theta.
\label{si2}
\end{equation}
For the Lifshitz case we can use the same parametrization. 
However, contrarily to the relativistic case, we no longer end up with a free theory. In fact, for $N=2$ the action (\ref{0.5}) reduces to 
\begin{equation}
S=\frac{1}{g}\int d^2x dt \left[\frac{1}{2}\partial_0\theta\partial_0\theta-\frac{a_1^2}{2}\partial_i\theta\partial_i\theta
-\frac{a_2^2}{2}\nabla^2\theta \nabla^2\theta-\frac{1}{2}(a_2^2+a_3+2a_4)(\partial_i\theta\partial_i\theta)^2 \right].
\label{si3}
\end{equation}
Thus we conclude that this interacting theory possesses anisotropic $z=2$ scale invariance at quantum level up to one loop when
the dimensional parameter is absent ($a_1=0$), since $\beta_g=\beta_{a_{2}}=0$ and then the quantum dilatation current is 
free of anomaly.

%%%%%%%%%%%%%%%%%%%%%%%%%%%%%%%%%%%%%%%%%%%

%%%%%%%%%%%%%%%%%%%%%%%%%%%%%%%%%%%%%%%%%%%%%%%%%%%%%%
\section{Supersymmetric Extension}\label{SS}

In this second part of the work we consider the supersymmetric extension of the Lifshitz nonlinear sigma model. 
In general supersymmetry improves the ultraviolet behavior of models what is expected to be favorable concerning 
Lorentz invariance restoration. 
We have essentially two general strategies for constructing Lifshitz-type supersymmetric models, according 
to the way the supersymmetry transformations are constructed. As they involve spacetime derivatives, 
we can construct an anisotropic theory invariant under a relativistic-like form of transformations or we can 
try to construct a nonrelativistic theory invariant under modified supersymmetric transformations.  
In the first case, the supercharges generating the supersymmetry transformations are still linear in derivatives, so that they satisfy the Leibniz rule as 
in the relativistic setting. On the other hand,  in the second case the 
supercharges will contain higher spatial derivatives, encoding themselves the anisotropic character of space and time.  
This case turns out to be equivalent to the supersymmetry arising from stochastic quantization
along the time direction (fictitious time). In contrast, the first possibility leads to a true 
spacetime supersymmetry. We will discuss both situations, starting with spacetime supersymmetry.

Let us start with the supersymmetric extension  in 2+1 dimensions from the superspace 
perspective. The superspace is constituted of bosonic coordinates $x^{0}$ and $x^{i}$, with $i=1,2$, and a pair of real Grassmannian coordinates $\theta_{\alpha}$, with $\alpha=1,2$. The generator of supersymmetry is constructed essentially as in the relativistic case, linear in 
space and time derivatives,
\begin{equation}
Q\equiv \frac{\partial}{\partial\bar\theta}+i \gamma^0\theta \partial_0+i a_1\gamma^i\theta\partial_i\equiv 
\frac{\partial}{\partial\bar\theta}+i \gamma^{\mu}\theta \tilde{\partial}_{\mu},~~~(\tilde{\partial}_{\mu}\equiv \partial_0,\,a_1\partial_i),
\label{1.6}
\end{equation}
where $\gamma^{\mu}$ are the Dirac matrices in a  2+1 dimensional representation,
chosen in terms of the Pauli matrices as $\gamma^0=\sigma_2$, $\gamma^1=i\sigma_1$ and 
$\gamma^2=i\sigma_3$. The conjugated Grassmann coordinate is $\bar \theta\equiv \theta^{T}\gamma^{0}$
and $a_1$ is a dimensionfull parameter ($[a_1]=1$ in mass units) to give the correct dimension for the supercharge since we have $[x^0]=2$ and $[x^i]=1$ in length units.  
The supercharge generates the following superspace translations
\begin{equation}
\delta x^0\equiv \bar\epsilon Q x^0=i\bar\epsilon \gamma^0 \theta,~~~
\delta x^i\equiv \bar\epsilon Q x^i=ia_1 \bar\epsilon \gamma^i \theta~~~\text{and}~~~\delta \theta\equiv \bar\epsilon Q \theta=\epsilon,
\label{1.7}
\end{equation}
where $\epsilon_{\alpha}$, $\alpha=1,2$, is a Grassmannian parameter of the transformation.

Consider the usual scalar superfield in 2+1 dimensions,
\begin{equation}
\Phi=\varphi+\bar\theta\psi+\frac{1}{2}\bar{\theta}\theta F,
\label{1.9}
\end{equation} 
where $\varphi$ is a real scalar field,  $\psi$ is a Majorana spinor field, and $F$ is an auxiliary bosonic degree of freedom.
The supersymmetry transformations for the components fields can be obtained from 
\begin{equation}
\delta\Phi\equiv \bar\epsilon Q\Phi.
\label{1.11}
\end{equation}
By using the properties $\bar{\theta}_{\alpha}\theta_{\beta}=i \theta _{2}\theta_{1}\delta_{\alpha\beta}=\frac{1}{2}\bar{\theta}\theta\delta_{\alpha\beta}$ and 
$\bar{\epsilon}\theta=\bar{\theta}\epsilon$, it follows that
\begin{eqnarray}
\delta\varphi&=& \bar\epsilon\psi,\nonumber\\
\delta \psi&=& -i\gamma^{\mu}\epsilon\tilde{\partial}_{\mu}\varphi+F\epsilon,\nonumber\\
\delta F&=& -i\bar{\epsilon} \gamma^{\mu}\tilde{\partial}_{\mu}\psi. 
\label{1.12}
\end{eqnarray}
Next we introduce the covariant superderivative
\begin{equation}
D\equiv \frac{\partial}{\partial\theta}-i \bar\theta\gamma^{\mu} \tilde{\partial}_{\mu},
\label{1.14}
\end{equation}
which anticommutes with the supercharge, i.e., $\{D,Q\}=0$. Thus any action in the superspace involving superfields and covariant 
derivative of superfields is manifestly supersymmetric.   

The supersymmetric nonlinear sigma model can be constructed from a set of scalar superfields $\Phi_A$, with $A=1,...,N$, 
subject to the constraint
\begin{equation}
\Phi_A\Phi_A=\frac{N}{2g},
\label{1.16}
\end{equation}
with $g$ being the coupling constant as in the nonsupersymmetric case. In terms of component fields the constraint
splits in
\begin{equation}
\varphi_A\varphi_A=\frac{N}{2g},~~~\psi_A\varphi_A=0,~~~\text{and}~~~\varphi_A F_A=\frac12 \bar\psi_A\psi_A,
\label{1.17}
\end{equation}
which are closed under (\ref{1.12}).
These constraints can be imposed via a superfield Lagrange multiplier
\begin{equation}
\Sigma=\sigma+\bar{\theta}\xi+\frac{1}{2}\bar{\theta}\theta\lambda,
\label{1.18}
\end{equation}
through the inclusion of $\Sigma(\Phi_A\Phi_A-N/2g)$ in the action.

So far there has been no reference to the anisotropic scaling except for the introduction of the parameter $a_1$, but this is artificial. 
The true anisotropic character must be encoded into the dynamics of the model, which
can be reached with a deformation of the relativistic case by the addition of 
$\Phi\nabla^2\Phi$. Note that it is manifestly supersymmetric since the operator $\nabla$ commutes with the supercharge,  $[\nabla,\,Q]=0$.
Thus we define the action for the anisotropic $z=2$ nonlinear sigma model as 
\begin{equation}
S=\frac12\int dt d^2x d^2\theta \left[\Phi_A \bar{D}{D}\Phi_A+ a_2\Phi_A \nabla^2\Phi_A   -
\Sigma (\Phi_A\Phi_A-\frac{N}{2g})\right],
\label{1.21}
\end{equation}
where $a_2$ is a dimensionless positive parameter. 

\begin{figure}[!h]
\centering
\includegraphics[scale=0.6]{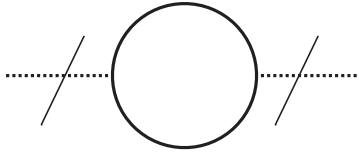}
\caption{Thick dotted lines represent $\Sigma$-field whereas continuous thick lines represent $\Phi$-field.}
\label{SuperGraph}
\end{figure}
Following the strategy of Sec. \ref{LN}, we will proceed with the $1/N$ expansion\footnote{The large-$N$ quantum behavior of the (2+1)-dimensional relativistic supersymmetric nonlinear sigma model is discussed in \cite{Koures}.}. 
By assuming that $\langle \Sigma\rangle=m^2$, we redefine the superfield $\Sigma$ such that the new field has a 
vanishing vacuum expectation value. This will generate a mass for the scalar superfield such that the superpropagator is
\begin{equation}
\langle\Phi_{A}(p,\theta_{1})\Phi_{B}(q,\theta_{2})\rangle=i\frac{\bar{D}D/2+a_2 {\bf p}^{2}+ m^2}{p_{0}^{2}-a_1^2{\bf p}^2- 
(a_2{\bf  p}^{2}+ m^2)^{2}} \delta_{AB}\delta^{3}(p-q)\bar{\delta}_{12},
\label{1.22}
\end{equation}
where $\bar{\delta}_{12}\equiv\delta(\bar{\theta}_{1}-\bar{\theta}_{2})\delta(\theta_{1}-\theta_{2}) $. By using this we see  that
 the supergraph of Fig. \ref{SuperGraph} furnishes the following contribution to the effective action
 \begin{eqnarray}
S_{\Sigma} &=& N \int \frac{dp_{0}}{2\pi}\frac{d^{2}p}{(2\pi)^{2}}\Sigma(p,\theta_1)  \Sigma(-p,\theta_2) \nonumber\\
&\times& \int \frac{dk_{0}}{2\pi}\frac{d^{2}k}{(2\pi)^{2}}d^2\theta_{1}d^{2}\theta_{2}\frac{[\bar{D}D/2+ a_2{\bf  k}^{2}+m^2]}
{\tilde{k}^2-(a_2{\bf k}^{2}+m^2)^{2}}\bar{\delta}_{12} \frac{[\bar{D}D/2+ a_2 ({\bf k+p})^{2}+ m^2]}{(\tilde{k}+\tilde{p})^{2}-
(a_2({\bf k+p})^{2}+m)^{2}}\bar{\delta}_{12},
\label{1.23}
 \end{eqnarray}
with $\tilde{k}^2\equiv k_0^2-a_1^2{\bf k}^2$.
It can be further simplified to 
\begin{equation}
S_{\Sigma} = N\int \frac{dp_{0}}{2\pi}\frac{d^{2}p}{(2\pi)^{2}}d^{2}\theta\Sigma(p,\theta) I_{1} [\bar{D}D/2+I_{2}I_{1}^{-1}+ 2m^2]  \Sigma(-p,\theta),
\label{1.24}
\end{equation}
where
\begin{eqnarray}
I_{1}&\equiv& \frac12\int \frac{dk_{0}}{2\pi}\frac{d^{2}k}{(2\pi)^{2}}\frac{1}{[\tilde{k}^{2}-(a_2{\bf k}^{2}+m^2)^{2}][(\tilde{k}+\tilde{p})^{2}-(a_2({\bf k+p})^{2}+m^2)^{2}]},\nonumber\\
I_{2}&\equiv& -\frac{a_2}2\int \frac{dk_{0}}{2\pi}\frac{d^{2}k}{(2\pi)^{2}}\frac{{\bf k}^{2}+({\bf k+p})^{2}}{[\tilde{k}^{2}-(a_2{\bf k}^{2}+m^2)^{2}][(\tilde{k}+\tilde{p})^{2}-(a_2({\bf k+p})^{2}+m^2)^{2}]}.
\label{1.25}
\end{eqnarray}
They have the asymptotic behaviors $I_{1}\approx |{\bf p}|^{-4}$ and $I_2\approx |{\bf p}|^{-2}$ for large momenta.
The effective action (\ref{1.24}) enable us to define the propagator for the $\Sigma$ superfield 
\begin{equation}
\langle\Sigma (p_{1},\theta_{1})\Sigma (p_{2},\theta_{2})\rangle =-\frac{1}{2 N}I_{1}^{-1}\frac{\bar{D}D-I_{2} I_{1}^{-1}- 2m^2}{\tilde{p}^{2}-(I_{2} I_{1}^{-1}+ 2m^2)^{2}}\delta^{3}(p_{1}-p_{2}) {\bar{\delta}}_{12}.
\label{1.26}
\end{equation}

%%%%%%%%%%%%%%%%%%%%%%%%%%%%%%%%%%%%%%%%%%

\subsection{Power Counting and Renormalizability}

From the large momentum behavior of the above propagators we can determine the superficial degree of divergence of the theory and discuss its renormalizability.
For a generic Feynman superdiagram $\mathcal{G}$, we  have
\begin{equation}
d(\mathcal{G})=4L-4n_{\Phi}+2(n_{\Phi}+n_{\Sigma}-L),
\label{1.27}
\end{equation}
where $n_\Phi$ and $n_\Sigma$ are the number of internal lines of the superfields $\Phi$ and $\Sigma$. The last term in the above expression comes from the numbers of $D$'s that can be converted into momenta. By using the Euler relation $L=n_{\Phi}+n_{\Sigma}-V+1$ and the
topological identities satisfied by the Feynman superdiagrams
$2n_{\Sigma}+N_{\Sigma}=V$ and $2n_{\Phi}+N_{\Phi}=2V$, we arrive at
\begin{equation}
 d(\mathcal{G})= 2 - 2 N_\Sigma,
\label{1.28}
 \end{equation} 
where $N_{\Sigma}$ is the number of external lines of the superfield $\Sigma$.
We see the improvement of the ultraviolet behavior compared to the nonsupersymmetric situation (\ref{4.4}). 
Here, for $N_\Sigma=0$, all superdiagrams involved are only quadratically divergent. 
The only nonreducible counterterms are that for $N_{\Phi}=2$,
\begin{equation}
\Phi \bar{D}D \Phi~~~\text{and}~~~\Phi \nabla^2 \Phi.
\label{1.29}
\end{equation} 
For an arbitrary 1PI vertex function with $N_\Phi \geq4$ the counterterms are always reducible to the above ones, 
for example,
\begin{equation}
(\Phi \bar{D}D \Phi) (\Phi\Phi)\cdots(\Phi\Phi)\sim (\Phi \bar{D}D \Phi)~~~\text{and}~~~ 
(\Phi \nabla^2 \Phi) (\Phi\Phi)\cdots(\Phi\Phi)\sim (\Phi \nabla^2 \Phi).
\label{1.30}
\end{equation} 
For $N_{\Sigma}=1$, all diagrams are logarithmically divergent. 
For $N_{\Phi}=0$, the counterterm is proportional to $\Sigma$. 
For $N_{\Phi}>0$, all diagrams are reducible. 
Thus we see that all necessary counterterms are of the form of the original Lagrangian and hence the 
model is renormalizable.

%%%%%%%%%%%%%%%%%%%%%%%%%%%%%%%%%%%%%
\subsection{Action in Components}

It is instructive to write the action (\ref{1.21}) in components to make evident 
how the anisotropic character in the superspace is passed to the ordinary spacetime. 
It reduces to
\begin{eqnarray}
S&=&\int dt d^2x\left[-\frac12 \varphi\tilde{\partial}^2\varphi+\frac{i}{2}\bar\psi\gamma^{\mu}\tilde{\partial}_{\mu}\psi+
\frac12 F^2
-a_2 F\nabla^2\varphi+\frac{a_2}{2}\bar\psi\nabla^2\psi\right.\nonumber\\
&+&\left. \sigma(F\varphi-\frac12\bar\psi\psi)-\bar\xi\psi\varphi+
\frac{\lambda}{2}(\varphi^2-\frac{N}{2g})\right].
\label{2.1}
\end{eqnarray}
The equation of motion of the auxiliary field $F_A$ reads
\begin{equation}
F_A=a_2\nabla^2\varphi_A-\sigma\varphi_A.
\label{2.2}
\end{equation}
Plugging it back into the Lagrangian we obtain
\begin{eqnarray}
S&=&\int dt d^2x\left[-\frac12 \varphi\tilde{\partial}^2\varphi-\frac{a_2^2}{2}\nabla^2\varphi\nabla^2\varphi+
\frac{i}{2}\bar\psi\gamma^{\mu}\tilde{\partial}_{\mu}\psi
+\frac{a_2}{2}\bar\psi\nabla^2\psi\right.\nonumber\\
&+&\left. a_2\sigma\varphi\nabla^2\varphi-\frac{1}{2}\sigma^2\varphi^2-\frac12\sigma\bar\psi\psi-\bar\xi\psi\varphi+
\frac{\lambda}{2}(\varphi^2-\frac{N}{2g})\right].
\label{2.3}
\end{eqnarray}
By integrating out the auxiliary field $F_A$, the $\sigma$-dependent part of the action becomes
\begin{equation}
\int \mathcal{D}\sigma \text{exp}\left[i\int dt d^2x\left(-\frac{N}{4g}\sigma^2+\sigma(a_2\varphi\nabla^2\varphi-\frac12\bar\psi\psi)\right)\right],
\label{2.4}
\end{equation}
where we have used the constraint $\varphi_A\varphi_A=N/2g$. The integration over this Lagrange multiplier
furnishes the quartic contributions to the action
\begin{equation}
\int dt d^2x \left[\frac{g a_2^2}{N}(\varphi\nabla^2\varphi)^2-
\frac{g a_2}{N}(\bar\psi\psi)(\varphi\nabla^2\varphi)+\frac{g}{4N} (\bar\psi\psi)^2  \right].
\label{2.5}
\end{equation}
%%%%%%%%%%%%%%%%%%%%%%%%%%%%%%%%%%%%%%%%%%%%%%

\subsection{Supersymmetry Breaking}
 
Let us consider the Lagrangian in the form (\ref{2.3}). It is convenient to redefine the fields $\sigma$ and $\lambda$  so that
\begin{equation}
\sigma\rightarrow \sigma+ m^{2}\qquad\mbox{and}\qquad \lambda\rightarrow \lambda+\lambda_{0}+ 2m^2\sigma + 2 m^{4},
\end{equation}
where $m^{2}$ and $\lambda_{0}+2 m^{4}$ are the vacuum expectation value of the $\sigma$ and $\lambda$ fields, respectively. After these shifts, the new fields $\sigma$ and $\lambda$ come to have vanishing vacuum expectation values. The action has the form
\begin{eqnarray}
S&=&\int dt d^2x\left[-\frac12 \varphi\tilde{\partial}^2\varphi-\frac{a_2^2}{2}\nabla^2\varphi\nabla^2\varphi+
\frac{i}{2}\bar\psi\gamma^{\mu}\tilde{\partial}_{\mu}\psi
+\frac{a_2}{2}\bar\psi\nabla^2\psi+ \frac{m^{4}}{2}\varphi^{2}\right.\nonumber\\
&+&\left. a_2\sigma\varphi\nabla^2\varphi+a_2 m^{2}\varphi\nabla^2\varphi-\frac{m^{2}}{2}\bar{\psi}\psi-\frac{1}{2}\sigma^2\varphi^2-\frac12\sigma\bar\psi\psi-\bar\xi\psi\varphi+
\frac{\lambda}{2}(\varphi^2-\frac{N}{2g})\right.\nonumber\\
&-&\frac{N}{2g}m^{2}\sigma+\left. \frac{\lambda_{0}}{2}\varphi^{2}\right]
\label{2.6}
\end{eqnarray}
From the quadratic terms we see that $m^{2}$ is the mass for the $\psi$ field. Thus, to keep the supersymmetry unbroken, we need to impose
$\lambda_{0}=-2m^{4}$. The free propagator for the $\psi$ field is then
\begin{eqnarray}
S_{AB}(p)&=& \frac{i\delta_{AB}}{\tilde{p}_{\mu}\gamma^{\mu}-(a_2 {\bf p}^{2}+m^{2})+i \delta_{AB}\epsilon}=
i\delta_{AB} \frac{\tilde{p}_{\mu}\gamma^{\mu}+(a_2 {\bf p}^{2}+m^{2})}{\tilde{p}^{2}-(a_2 {\bf p}^{2}+m^{2})^2+i \epsilon}.
\end{eqnarray}
The propagator for the scalar field is
\begin{equation}
\Delta_{AB}(p)=  \frac{i\delta_{AB}}{\tilde{p}^{2}-(a_2{\bf p}^{2}+m^{2})^{2}}.
\end{equation}
The conditions that $\langle\sigma\rangle=0$ and $\langle\lambda\rangle=0$ yield to two gap equations, 
which for $\lambda_0=-2m^4$ reduces to the same one, 
\begin{equation}
\int \frac{dk_{0}}{2\pi}\frac{d^{2}k}{(2\pi)^{2} }\frac{i}{\tilde{k}^{2}-(a_2{\bf k}^{2}+m^{2})^{2}}= \frac{1}{2g}.
\end{equation}
If $\lambda_{0}\not = -2 m^{4}$ supersymmetry is broken and the gap equations become incompatible. By following 
the same steps as in the nonsupersymmetric case we obtain the $\beta$-function
\begin{equation}
\beta_{g_R}=-\frac{2}{\pi}\frac{\mu^2}{(a_1^2+2a_2\mu^2)}g_R^2,
\label{2.7}
\end{equation}
which differs from (\ref{3.6}) by a factor of 2. Thus, by the same reasoning as before, this $\beta$-function is also compatible 
with low-energy Lorentz symmetry restoration in the strong form, since it is stable even when $a_2\rightarrow 0$.

Although it is hard to promote any further renormalization group calculation, it is worth to list the propagators of the auxiliary fields since
we can see a typical canceling of ultraviolet divergences in supersymmetric theories. 

$\bullet$ Auxiliary field $\lambda$: $\Delta_{\lambda}= - 1/F_{1}$, where
\begin{equation}
F_{1}(p)= \frac{N}{2}\int \frac{dk_{0}}{2\pi}\frac{d^{2}k}{(2\pi)^{2}}
\frac{1}{[(\tilde{k}+\tilde{p})^{2}-(a_2({\bf k}+{\bf p})^2+m^2)^{2}][\tilde{k}^{2}-(a_2{\bf k}+m^2)^{2}]},
\end{equation}
which is finite. 

$\bullet$ Auxiliary field $\sigma$: $\Delta_{\sigma}=-1/F_{2}$, where
\begin{eqnarray}
F_{2}(p)&=& N \int\frac{dk_{0}}{2\pi}\frac{d^{2}k}{(2\pi)^{2}}\frac{1}{\tilde{k}^{2}-(a_2{\bf k}^2+m)^{2}}\nonumber\\&+&
Na_2^{2}\int\frac{dk_{0}}{2\pi}\frac{d^{2}k}{(2\pi)^{2}}\frac{( {\bf k}^{2})^{2}+ {\bf k}^{2}({\bf k+p})^{2}}{[(\tilde{k}+\tilde{p})^{2}-(a_2({\bf k}+{\bf p})^2+m^2)^{2}][\tilde{k}^{2}-(a_2{\bf k}+m^2)^{2}]}.
\end{eqnarray}
While these integrals are individually logarithmically divergent $F_2$ turns out to be finite, as can be seen after integration on $k_0$. 

$\bullet$ Auxiliary field $\xi$: $S_{\xi}= -1/F_{3}$, where
\begin{equation}
F_{3}(p)= -N \int \frac{dk_{0}}{2\pi}\frac{d^{2}k}{(2\pi)^{2}}\frac{\tilde{k}_{\mu}\gamma^{\mu}+a_2{\bf k}^2+m^2}
{[(\tilde{k}+\tilde{p})^{2}-(a_2({\bf k}+{\bf p})^2+m^2)^{2}][\tilde{k}^{2}-(a_2{\bf k}+m^2)^{2}]},
\end{equation}
which is finite. 

As it happened in the nonlinear sigma model without supersymmetry we are not able to evaluate the renormalization group flow of parameter 
$a_2$. Nevertheless, the supersymmetric extension exhibits the same pattern concerning Lorentz restoration at low energies as the 
nonsupersymmetric counterpart, since the beta function (\ref{2.7}) is well behaved when $a_2\rightarrow 0$.
Furthermore, the supersymmetry brings advantages as the requirement of a fewer number of parameters (the only Lorentz violating 
parameter is $a_2$) and the improvement of the ultraviolet behavior.

%%%%%%%%%%%%%%%%%%%%%%%%%%%%%%%%%%%%%%%%%%%%%%%%%%%%%%%%

\section{Modified Supersymmetry Transformations}

In the previous section the nonrelativistic supersymmetry was obtained by the inclusion of 
supersymmetric but non Lorentz invariant term $\Phi\nabla^2\Phi$ into the action. The supersymmetry transformations (\ref{1.12})
are essentially the same as the relativistic ones, generated by supercharges which are linear in the derivatives, such that the 
construction of the superspace is straightforward.  

It is natural to investigate the possibility of incorporating the anisotropic character of space and time into 
the transformations of supersymmetry themselves. 
To this end we are tempted to introduce supercharges of the form
\begin{equation}
Q\equiv \frac{\partial}{\partial\bar{\theta}}+i\gamma^0\theta\partial_0-\vartheta\theta\nabla^2~~~\text{and}~~~
\bar{Q}\equiv\frac{\partial}{\partial{\theta}}+i\bar{\theta}\gamma^0\partial_0-\vartheta\bar{\theta}\nabla^2,
\label{4.1a}
\end{equation}
where the dimensionless parameter $\vartheta$ is free for the moment. 
An immediate problem with these supercharges is that they do not satisfy the Leibniz rule such that it is unclear the  
construction of manifestly supersymmetric actions. 
Furthermore, they do not generate spatial translations,  
\begin{equation}
\delta x^0\equiv \bar\epsilon Q x^0=i\bar\epsilon \gamma^0 \theta~~~\text{and}~~~
\delta x^i\equiv \bar\epsilon Q x^i=0
\label{4.2a}
\end{equation}
and 
\begin{equation}
\delta \theta\equiv \bar\epsilon Q \theta=\epsilon.
\label{4.3a}
\end{equation}
Although these weird points, the above supercharges generate transformations with the desirable 
features, i.e., they encode the anisotropic character of space and time. In fact, from $\delta\Phi\equiv \bar\epsilon Q\Phi$, 
where $\Phi$ is the superfield of (\ref{1.9}), we obtain
\begin{eqnarray}
\delta\varphi&=& \bar\epsilon\psi,\nonumber\\
\delta \psi&=& -i\gamma^0\epsilon \partial_0\varphi+F\epsilon-\vartheta\nabla^2\varphi\epsilon,\nonumber\\
\delta F&=& -i\bar{\epsilon}  \gamma^0\partial_0\psi+\vartheta\bar{\epsilon}\nabla^2\psi.
\label{4.6a}
\end{eqnarray}
Even without a superspace formulation we should be able to proceed 
with the supersymmetry in components. The peculiarity in the case of a nonlinear sigma model is the existence of the constraints which need to 
be closed under supersymmetry transformations in order to produce a consistent theory. By starting with the bosonic constraint 
\begin{equation}
\varphi_A\varphi_A=\frac{N}{2g}
\label{4.8a}
\end{equation}
and performing successive transformations we end up with
\begin{equation}
\psi_A\varphi_A=0~~~\text{and}~~~2F_A\varphi_A-2\vartheta\varphi_A\nabla^2\varphi_A-\bar{\psi}_A\psi_A=0,
\label{4.9a}
\end{equation}
which, as required, are closed. 
We see that there is an additional contribution compared to (\ref{1.17}), 
showing that the constraints must be modified to be compatible with (\ref{4.6a}). 
Of course, these constraints cannot be obtained from the superfield expression 
$\Phi_A\Phi_A=\frac{N}{2g}$. 
However, we can eliminate the higher derivative $\vartheta$-terms in the expressions in (\ref{4.6a}) and (\ref{4.9a})
just by means of a redefinition of the auxiliary field $F\rightarrow F+\vartheta \nabla^2\varphi$, such that the new 
transformations involve only time derivatives 
\begin{eqnarray}
\delta\varphi&=& \bar\epsilon\psi,\nonumber\\
\delta \psi&=& -i\gamma^0\epsilon \partial_0\varphi+F\epsilon,\nonumber\\
\delta F&=& -i\bar{\epsilon}  \gamma^0\partial_0\psi,
\label{4.6aa}
\end{eqnarray}
while the constraints recover the relations in (\ref{1.17}). 
In its turn, these new transformations are generated by the supercharges
\begin{equation}
Q_0= \frac{\partial}{\partial\bar{\theta}}+i\gamma^0\theta\partial_0~~~\text{and}~~~
\bar{Q}_0=\frac{\partial}{\partial{\theta}}+i\bar{\theta}\gamma^0\partial_0,
\label{4.12a}
\end{equation}
which are now linear operators in the derivatives. 
Thus we can obtain a superspace formulation by introducing the 
covariant derivatives,
\begin{equation}
D_0= \frac{\partial}{\partial\theta}-i \bar\theta\gamma^0 \partial_0~~~\text{and}~~~
\bar{D}_0= \frac{\partial}{\partial\bar\theta}-i \gamma^0\theta \partial_0,
\label{4.13a}
\end{equation}
such that $\{D_0,Q_0\}=0$. Any action involving superfields and covariant derivatives of superfield is manifestly supersymmetric.
The kinetic term constructed from (\ref{4.13a}) gives
\begin{equation}
\mathcal{L}_{\text{kin}}=\frac{1}{2}\int d^2\theta \Phi\bar{D}_0D_0\Phi=-\frac12 \varphi{\partial}_0^2\varphi+\frac{i}{2}\bar\psi\gamma^{0}{\partial}_{0}\psi+\frac12 F^2,
\label{4.14a}
\end{equation}
where we are omitting the internal index $A$.
As in the previous section, in order to produce higher spatial derivative terms for $\varphi$ and $\psi$ we need to include a term
$\Phi\nabla^2\Phi$ in the Lagrangian,
\begin{equation}
\mathcal{L}=\frac{1}{2}\int d^2\theta \left( \Phi\bar{D}_0D_0\Phi +a_2 \Phi\nabla^2\Phi-\Sigma(\Phi^2-\frac{N}{2g})\right).
\label{4.15a}
\end{equation}
So it is clear that this formulation is equivalent to the case $a_1=0$ in the construction of Sec. \ref{SS}.
This Lagrangian arises in the context of stochastic quantization of the nonlinear sigma model in 
a 2-dimensional Euclidean spacetime \cite{Brunelli}. Following the stochastic quantization 
recipe, we need to introduce an additional time coordinate in the system, called fictitious time, 
such that evolution along it is governed by a Langevin equation.
The fictitious time is then properly eliminated after the computation of the correlation functions, producing a
2-dimensional but quantum theory. 
In the construction of Lifshitz theories we just keep the fictitious time and interpret it as a physical 
time coordinate. In this sense, the stochastic quantization gives a prescription for obtaining $(d+1)$-dimensional 
Lifshitz-type theories from $d$-dimensional relativistic ones \cite{Orlando}.

%%%%%%%%%%%%%%%%%%%%%%%%%%%%%%%%%%%%%%%%%%%%%%%%%%%%%%%%%%%%
\section{Final Considerations}

The search for ultraviolet complete theories have attracted attention to Lifshitz extensions of relativistic theories. Adopting this procedure, 
 we can turn nonrenormalizable theories in renormalizable ones at the price of Lorentz symmetry. 
The most expressive case was pointed out by Horava \cite{Horava}, who  proposed a power counting renormalizable 
gravitational theory in 3+1 dimensions with $z=3$.  
Lifshitz type theories are interesting in their own right offering a new framework to
investigate several field theoretical properties. Furthermore, they are quite useful 
in the description of quantum condensed matter systems with dynamical scaling weighted by the critical exponent $z$ In fact, anisotropic scaling was originally proposed in the study of Lifshitz points.

Naive dimensional analysis indicates that higher spatial derivative terms are suppressed at low energies when compared 
to lower derivative operators (which in the weighted power counting are relevant operators). However, some
perturbative renormalization group calculations \cite{Iengo,Gomes} show that the resurgence of  Lorentz invariance at low energy 
depends strongly on a fine-tuning of the Lorentz violating parameters, in the sense that they cannot be 
arbitrarily small since the renormalization group functions become ill-defined  as a reflex of the nonrenormalizability. 
We call this situation of Lorentz restoration in the weak form and it is what often happens within the perturbative framework of 
strictly renormalizable Lifshitz-type models.  

Lifshitz nonlinear sigma models admit large-$N$ expansion which 
better incorporate Lorentz invariance issue at low energies than the perturbative expansion. 
We conclude this by computing the renormalization group flow of the coupling constant, which is sensitive to the 
relevant operator $a_1^2(\partial_i\varphi)^2$ making it well defined even when $a_2$ is arbitrarily small. 
Unfortunately, loop calculations with anisotropic propagators are hard to be thoroughly performed preventing us 
to determine the flow of the Lorentz violating parameter.  
In view of these difficulties, we performed a perturbative study of the model to 
gain some more information concerning the running of the parameters. 
We obtained that the parameter $a_2$ does not run up to the considered order, i.e., $\beta_{a_2}=0$. 
We interpret this as a positive signal  concerning Lorentz restoration. Such a behavior indicates that there is no dangerous 
behavior of $a_2$ under renormalization group flow that otherwise would require some fine-tuning to be tamed.  
We showed also that the perturbative $\beta_g$ is not sensitive to the presence of the relevant operator 
$a_1^2(\partial_i\varphi)^2$, contrarily to the result of the large-$N$ behavior.

In the course of these studies we come across some interesting questions. 
In the perturbative context, we find a potential interacting theory possessing anisotropic scale 
invariance at quantum level. This type of model can be useful in condensed matter systems to describe a universality class
of quantum phase transitions with $z=2$ in 2+1 dimensions.
Some technical issues were also addressed in the beginning the work, namely, the construction of generalized sigma models. 
In the relativistic context, such constructions are interesting since they 
put together a class of models exhibiting common features such as asymptotic freedom and  
an infinite number of non-local classical conserved charges \cite{Brezin,Hikami}. 
Our intention was to describe a more general framework for constructing 
Lifshitz-type models. For example, by choosing $q\in U(N)$ we obtain an anisotropic $z=2$ version 
of the $CP(N)$ model \cite{Das}.

The supersymmetric extension of the model exhibits the same general pattern 
concerning low-energy Lorentz restoration as the nonsupersymmetric counterpart,
corresponding to the restoration in the strong form.
The beta function of the coupling constant is well behaved when $a_2\rightarrow 0$ and  
the advantage is that it is the only one Lorentz violating parameter required by renormalizability. 
This is a basic consequence of the improvement of ultraviolet behavior enforced by supersymmetry.

The general construction of Lifshitz-type supersymmetric theories is challenging. As it is known, 
the canceling of divergences becomes automatic when formulated in superspace. On the other hand, the 
construction of superspace in Lifshitz theories is problematic. This is so because the natural candidates for 
supercharges generating anisotropic transformations involve higher spatial derivatives 
such that they no longer satisfy the Leibniz rule. Hence it is hard to write down manifestly supersymmetric actions and 
even the notion of chiral and anti-chiral superfields is obscure. 

The case considered here does not suffer from these problems since we keep the supersymmetry 
transformations unchanged compared to the relativistic ones. The anisotropic character of space and 
time is induced by the inclusion of a supersymmetric but non Lorentz invariant term $\Phi\nabla^2\Phi$.
When trying to discuss modifications in the supercharges, we end up with a theory 
with the supersymmetry only in the time direction, being equivalent to the stochastic quantization 
supersymmetry arising in the fictitious time. 

Further studies are necessary to ultimately establish the Lorentz restoration in the strong form. 
This involves the determination of the renormalization group flow 
of all parameters of the theory, but certainly it would require to go beyond our analytic computations, 
being necessary to resort to numerical methods. 
In addition, the future direction is to extend these investigations to other Lifshitz-type theories which admit large-$N$ expansion. 
Of course, their usefulness in high-energy physics is highlighted whenever we are able to identify 
Lorentz invariant sectors in their spectrum.

%%%%%%%%%%%%%%%%%%%%%%%%%%%%%%%%%%%%%%%%%%%%%%%%%%%%

%%%%%%%%%%%%%%%%%%%%%%%%%%%%%%%%%%%%%%%%%%%%%%
\section{Acknowledgments}

We would like to thank Carlos Hernaski for very helpful discussions. 
This work was partially supported by  Funda\c{c}\~ao de Amparo a Pesquisa do Estado de S\~ao Paulo (FAPESP) and
Conselho Nacional de Desenvolvimento Cient\'ifico e Tecnol\'ogico (CNPq).

%%%%%%%%%%%%%%%%%%%%%%%%%%%%%%%%%%%%%%%%%%%%%%%%%%%%%
\appendix

%%%%%%%%%%%%%%%%%%%%%%%%%%%%%%%%%%%%%%%%%%%%%%%%%%%%%%%%

\section{Perturbative Renormalization of the $z=2$ Lifshitz Nonlinear Sigma Model}\label{BB}

The perturbative renormalizability in $d=2$ can be established by using the Ward identity coming from the constraint, following 
a similar strategy as in the relativistic case \cite{Amit}.
Consider the generating functional in the presence of sources $H(x)$ and $J_A(x)$, with $A\neq 1$,
\begin{equation}
Z[J_A,H]=\int \mathcal{D}\pi \exp\left[ iS[\phi,\pi]+i\int d^2xdt(H\phi +J_A \pi_A)\right],
\label{A1}
\end{equation}
where $\mathcal{D}\pi\equiv \prod_A\mathcal{D}\pi_A \mathcal{J}$ and the Jacobian $\mathcal{J}$ is to ensure the $O(N)$ invariance 
of the integration measure\footnote{This Jacobian can be reduced to the unit when dimensionally regularized \cite{Amit}.}.
The action $S$ is invariant under the nonlinear transformations
\begin{equation}
\delta \pi_A= \left(\frac{N}{2g}-\pi^2\right)^{\frac{1}{2}}\omega_A~~~\text{and}~~~\delta\phi=-\omega_A \pi_A,
\label{A2}
\end{equation}
where $\omega_A$ are the parameter of the transformations. This symmetry leads to the Ward identity 
\begin{equation}
\int d^2xdt\left(J_A(x) \frac{\delta }{\delta H(x)}-H(x)\frac{\delta}{\delta J_A(x)}\right)Z[J_A,H]=0.
\label{A3}
\end{equation}
A similar expression follows for the generating functional of the connected Green functions $W$, $Z=e^{iW}$.
After a Legendre transformation we can write the Ward identity for the effective action~$\Gamma$,
\begin{equation}
\int d^2xdt\left( \frac{\delta \Gamma }{\delta H(x)} \frac{\delta \Gamma }{\delta \bar{\pi}_A(x)}+H(x) \bar{\pi}_A(x)\right)=0,
\label{A4}
\end{equation}
where $\bar{\pi}_A(x)$ is the classical field given by
\begin{equation}
\bar{\pi}_A(x)=\frac{\delta W[J_A,H]}{\delta J_A}.
\label{A5}
\end{equation}
The divergent part of $\Gamma$,  which will be denote by $\Gamma_{div}$, must also to satisfy the above identity.
Dimensional analysis is useful to proceed. Note that $[H]=4$ in mass units. We can expand the divergent part of the effective action 
in powers of $H$. On dimensional grounds it follows,
\begin{equation}
\Gamma_{div}=\int d^2xdt\left[ \Gamma^{(0)}(\bar\pi,\partial_{0}\bar\pi,\partial_i\bar\pi,\partial_i\partial_j\bar\pi)+H (x)\Gamma^{(1)}(\bar\pi)\right].
\label{A6}
\end{equation}
We can divide $\Gamma^{(0)}$ in three independent parts:
\begin{equation}
 \Gamma^{(0)}(\bar\pi,\partial_{0}\bar\pi,\partial_i\bar\pi,\nabla^2\bar\pi)=\Gamma_A^{(0)}(\bar\pi,\partial_0\bar\pi)+
 \Gamma_B^{(0)}(\bar\pi,\partial_i\bar\pi)+
 \Gamma_C^{(0)}(\bar\pi,\partial_i\partial_j\bar\pi).
 \label{A7}
\end{equation}
Note that there is no mixing of time and spatial derivatives. The only dimension four possible operator containing such a mixing would be the form
$\partial_0\bar\pi\partial_i\partial_i\bar\pi f(\bar\pi)$ which is forbidden by the symmetry $x^0\rightarrow -x^0$.

Plugging these expressions into the Ward identity (\ref{A4}) we have,
\begin{equation}
\int  d^2x dt \left[\Gamma^{(1)}\left(\frac{\delta A^{(0)}}{\delta\bar\pi_A}+H\frac{\partial \Gamma^{(1)}}{\partial\bar\pi_A}\right)+H\bar{\pi}_A\right]=0,
\label{A8}
\end{equation}
where $A^{(0)}\equiv \int d^2x dt \Gamma^{(0)}$. From this we see that,
\begin{equation}
\int d^2x dt\, \Gamma^{(1)}\frac{\delta A^{(0)}}{\delta\bar\pi_A}=0~~~\text{and}~~~
\Gamma^{(1)}\frac{\partial \Gamma^{(1)}}{\partial\bar\pi_A}+\bar\pi_A=0.
\label{A9}
\end{equation}
The solution of the second condition is
\begin{equation}
\Gamma^{(1)}(\bar{\pi})=(b-\bar\pi^2)^{\frac{1}{2}},
\label{A10}
\end{equation}
where $b$ is an integration constant.
The first condition implies that
\begin{equation}
\int d^2x dt\, \Gamma^{(1)}\frac{\delta A_A^{(0)}}{\delta\bar\pi_A}=0,~~~
 \int d^2x dt\, \Gamma^{(1)}\frac{\delta A_B^{(0)}}{\delta\bar\pi_A}=0,~~~
\int d^2x dt\, \Gamma^{(1)}\frac{\delta A_C^{(0)}}{\delta\bar\pi_A}=0,
\label{A11}
\end{equation}
with $A_I^{(0)}\equiv \int d^2x dt \,\Gamma_I^{(0)}$, $I=A,B,C$.
These conditions correspond to the part with $H=0$ and hence must be invariant under the 
full $O(N)$ group. Invariant terms with two time derivatives can be constructed from $(\partial_0\bar\varphi_A)^2$, 
where $\bar{\varphi}_A\equiv(\sqrt{c-\bar\pi^2},\bar{\pi}_A)$.  
Thus, 
\begin{equation}
A_A^{(0)}=\int d^2x dt c_1\left(\partial_0\sqrt{c-\bar\pi^2}\partial_0\sqrt{c-\bar\pi^2}+\partial_0\bar\pi\partial_0\bar\pi\right),
\label{A12}
\end{equation}
with $c_1$ being an arbitrary constant.
With the same reasoning, from $(\partial_i\bar\varphi_A)^2$ we write
\begin{equation}
A_B^{(0)}=\int d^2x dt c_2\left(\partial_i\sqrt{c-\bar\pi^2}\partial_i\sqrt{c-\bar\pi^2}+\partial_i\bar\pi\partial_i\bar\pi\right).
\label{A13}
\end{equation}
With four spatial derivatives, we have three possibilities $(\nabla^2\bar\varphi)^2$, $(\bar\varphi\nabla^2\bar\varphi)^2$ and 
$(\bar\varphi\partial_i\partial_j\bar\varphi)^2$, 
\begin{eqnarray}
A_C^{(0)}&=&\int d^2x dt  \left[c_3\left( (\nabla^2\sqrt{c-\bar\pi^2})^2+(\nabla^2\bar\pi)^2\right)+
c_4\left(\sqrt{c-\bar\pi^2}\nabla^2\sqrt{c-\bar\pi^2}+\bar\pi\nabla^2\bar\pi\right)^2\right.\nonumber\\&+&\left.  
c_5\left(\sqrt{c-\bar\pi^2}\partial_i\partial_j\sqrt{c-\bar\pi^2}+\bar\pi\partial_i\partial_j\bar\pi\right)^2\right].
\label{A14}
\end{eqnarray}
The constant $c$ defines the new $O(N)$ transformation
\begin{equation}
\delta\bar\pi_A=(c-\bar\pi^2)^{\frac{1}{2}}\omega_A.
\label{A15}
\end{equation}
All conditions in (\ref{A11}) are simultaneously satisfied 
for $c=b$.  Thus, the divergences contained in $c,c_1,c_2,c_3,c_4,c_5$ can be absorbed in the 
redefinition of the parameters $g,a_1,a_2,a_3,a_4$ and $Z$, as in (\ref{0.8c}), ensuring the renormalizability of the model. 

%%%%%%%%%%%%%%%%%%%%%%%%%%%%%%%%%%%%%%%%%%%%%%%%%%%%%%%

%%%%%%%%%%%%%%%%%%%%%%%%%%%%%%%%%%%%%%%%%%
\end{document}